\documentclass[showkeys,12pt,onecolumn]{article}

\usepackage{arxiv}
\usepackage[sorting = none, backend = bibtex, style=numeric-comp]{biblatex}
\usepackage[english]{babel}
\usepackage[utf8]{inputenc}
\usepackage[colorinlistoftodos, color=green!40, prependcaption]{todonotes}
\usepackage[colorlinks=true,linkcolor=blue,urlcolor=blue,citecolor=blue]{hyperref}
\usepackage{mathrsfs}
\usepackage{amsmath}
\usepackage{amsfonts}
\usepackage{amssymb}
\usepackage{amsthm}
\usepackage{slashed}
\usepackage{mathtools}
\usepackage{xcolor}
\usepackage{enumitem}
\usepackage[capitalise]{cleveref}
\usepackage[normalem]{ulem}
\usepackage{lipsum}

\makeatletter
\newcommand{\mybf}[1]{%
    \begingroup
    \def\@tempa{#1}%
    \ifx\@tempa\@empty
        \textbf{#1}%
    \else
        \edef\@tempa{\noexpand\in@{#1}{abcdefghijklmnopqrstuvwxyzABCDEFGHIJKLMNOPQRSTUVWXYZ}}%
        \@tempa
        \ifin@
            \mathbf{#1}%
        \else
            \pmb{#1}%
        \fi
    \fi
    \endgroup
}
\makeatother

\begin{document}

\title{Thermal fluid closures and pressure anisotropies in numerical simulations of plasma wakefield acceleration}

\author{Daniele Simeoni$^{1,*}$
\And Andrea Renato Rossi$^{2}$
\And Gianmarco Parise$^{3,4}$
\And Fabio Guglietta$^{1}$
\And Mauro Sbragaglia$^{1}$
\and \\
$^{1}$Department of Physics \& INFN, Tor Vergata University of Rome, Via della Ricerca Scientifica 1, 00133, Rome, Italy
\and
$^{2}$INFN, Section of Milan, via Celoria 16, 20133, Milan, Italy
\and
$^{3}$Department of Physics, Tor Vergata University of Rome, Via della Ricerca Scientifica 1, 00133, Rome, Italy
\and
$^{4}$INFN, Laboratori Nazionali di Frascati, Via Enrico Fermi 54, 00044, Frascati, Italy
\and
$^{*}$\texttt{daniele.simeoni@roma2.infn.it}
}

\date{\today} 

\maketitle{} 

\begin{abstract}
We investigate the dynamics of plasma-based acceleration processes with collisionless particle dynamics and non-negligible thermal effects. We aim at assessing the applicability of fluid-like models, obtained by suitable closure assumptions applied to the relativistic kinetic equations, thus not suffering of statistical noise, even in presence of a finite temperature. The work here presented focuses on the characterization of pressure anisotropies, which crucially depend on the adopted closure scheme, and hence are useful to discern the appropriate thermal fluid model. To this aim, simulation results of spatially resolved fluid models with different thermal closure assumptions are compared with the results of particle-in-cell (PIC) simulations at changing temperature and amplitude of plasma oscillations.
\end{abstract}
\maketitle

\section{Introduction} \label{sec:intro}

Plasma wakefield acceleration (PWFA) is a promising new concept in the context of accelerator physics~\cite{shiltsev-2021, ferrario-2021, gourlay-2022}. In this process, a bunch of charged relativistic particles (driver) enters a neutral plasma and creates a perturbation in the form of oscillations of plasma electrons; the resulting electromagnetic fields produce intense accelerating forces that are orders of magnitude larger than those obtained with conventional accelerating technologies~\cite{modena-1995, blumenfeld-2007, litos-2016}. The physics of plasma-based acceleration processes is extremely rich and complex~\cite{tajima-1979,chen-1985,joshi-2007,esarey-2009,hogan-2016}: the driver moves at relativistic velocities, and it interacts with the plasma particles via electromagnetic forces: the early stages of the dynamics take place on time-scales much smaller than the characteristic time of inter-particle collisions, thus precluding thermalization to local equilibrium for plasma particles. The resulting dynamic is well represented by the relativistic Maxwell-Vlasov kinetic equations~\cite{vlasov-1961, cercignani-2002}.\\ 
Due to the inherent complexity of the physics involved, numerical simulations are important tools to support the design of novel PWFA experimental configurations. Particle-in-cell (PIC) methods are the reference tool in the community~\cite{shalaby-2017,terzani-2020,arber-2015,lehe-2016,fonseca-2002,bussmann-2013,derouillat-2018}. In PIC simulations, the relativistic Vlasov-Maxwell equations are solved via particle dynamics at the microscopic level~\cite{dawson-1983, birdsall-2018}. While this provides the most accurate description of the physics, it also requires a careful balance between increasing computational costs and low statistical noise, which both scale with the number of simulated particles~\cite{kesting-2015, birdsall-2018, hockney-2021}. On the other hand, there has been some effort in the community to design coarse-grained models based on the fluid description of plasma~\cite{muscato-1993}: instead of particle distribution functions, these methods provide a continuum description based on fields such as particle density and velocity. Fluid models result from suitable closure assumptions applied to the kinetic equations~\cite{toepfer-1971,newcomb-1982,amendt-1985,schroeder-2005, schroeder-2009, schroeder-2010}, and therefore can not be used to describe the whole richness emerging from kinetic models. Nevertheless, an accurate determination of the limits on their applicability can help in designing efficient tools of analysis for PWFA as a valid alternative to PIC simulations~\cite{massimo-2016-a,marocchino-2016}. \\
Fluid models have been traditionally developed in the cold limit, i.e., by neglecting thermal effects in the plasma dynamics~\cite{akhiezer-1956}. Thermal effects are known to provide a regularization of the singularity emerging in the wakefield structure close to wake-breaking~\cite{katsouleas-1988, rosenzweig-1988, schroeder-2005, jain-2015}; furthermore, recent research has prompted renewed interest in considering plasma acceleration processes with non negligible thermal effects since they may become relevant for high repetition rate operation~\cite{gholizadeh-2011,gilljohann-2019,darcy-2022}, have significant impact in the understanding of ion channel formations~\cite{zgadzaj-2020,khudiakov-2022}, improve beam quality in positron based accelerators~\cite{silva-2021,diederichs-2023,diederichs-2023-b,cao-2024} and also help in the development of new PWFA diagnostics~\cite{lee-2024}. \\
Developing fluid models for PWFA involving collisionless particle dynamics and including non-negligible thermal effects is a non-trivial task and requires the application of suitable closure schemes to the relativistic kinetic equations~\cite{toepfer-1971, newcomb-1982, amendt-1985, newcomb-1986-b, siambis-1987, pennisi-1991,muscato-1993}. Applying a closure scheme that is based on the assumption that the distribution function is close to a local equilibrium (hereafter named "local equilibrium closure" , LEC) seems inappropriate, at least in the early stages of plasma dynamics where particles do not collide and the local particle distribution function cannot relax towards a local equilibrium. However, if the plasma is initially in equilibrium and at rest, after the perturbation induced by the driver there may be features in the dynamics that may be qualitatively captured by LEC when the perturbation is small (linear regimes); other applications of LEC could be related to late stage dynamics, after the wake has mixed and broken~\cite{khudiakov-2022}. In fact, it is known \cite{schmidt-1979} that nonlinear collective forces have the same effect as collisions in thermalizing a particle distribution. Hence, one could think of using LEC for some numerical investigations, especially thanks to the fact that the number of fluid equations is limited, i.e., one has to consider in this case only mass and momentum conservation~\cite{toepfer-1971,cercignani-2002,rezzolla-2013}. Other closure schemes~\cite{amendt-1985, newcomb-1982, siambis-1987,schroeder-2005, schroeder-2009, schroeder-2010} can also be derived by considering centered moments equations obtained from the Vlasov equation and neglecting moments higher than the second without any assumption on the local equilibrium structure. This closure scheme (hereafter named "warm closure", WARMC) reproduces fluid equations possessing a pressure field that is inherently anisotropic~\cite{shadwick-2004,shadwick-2005}. With respect to the LEC, the WARMC involves a larger number of fluid equations  and is nominally valid only for small thermal spreads i.e. momentum distribution variances. Both fluid closures (LEC and WARMC) coincide in the cold limit, but they are expected to deliver different results at finite temperatures. So far, a quantitative assessment of the validity of spatially resolved thermal fluid closures in numerical simulations of PWFA is missing. This paper takes a step further in filling this gap. In more detail, we will focus here on a characterization of pressure anisotropies in thermal plasmas~\cite{shadwick-2004,shadwick-2005,khalilzadeh-2015}, specifically assessing the ability of fluid closures in describing this particular feature of PWFA experiments by comparing them against PIC simulations, the most reliable benchmark aside from conducting actual experiments. This work builds on the work presented in~\cite{shadwick-2004,shadwick-2005,khalilzadeh-2015}, and expands it by presenting fully spatially resolved results.
The paper is organized as follows. In Sec.~\ref{sec:relativistic}, the relativistic kinetic equations for collisionless plasma are recalled and the main ingredients for the two closure schemes (LEC and WARMC) are given. The problem setup and the numerical schemes are described in Sec.~\ref{sec:setup}. Results are described in Sec.~\ref{sec:results} and conclusions will follow in Sec.~\ref{sec:conclusions}.

\section{Warm Fluid Theories for a collisionless plasma}\label{sec:relativistic}
The basic step in the construction of a fluid theory for warm plasmas is provided by the relativistic Vlasov equation~\cite{vlasov-1961, cercignani-2002},
\begin{align}\label{eq:vlasov-eq}
p^{\alpha} \frac{\partial f}{\partial x^\alpha} - \frac{e}{c} F^{\alpha\beta} p_{\beta} \frac{\partial f}{\partial p^\alpha} = 0    \; ,
\end{align}
that details the time evolution of a gas of electrons with mass $m_e$ and charge $-e$, space-time coordinates $x^{\alpha}=(ct, \mybf{x})$ and relativistic kinetic momentum $p^{\alpha}=(p^0, \mybf{p})$, via the distribution function $f=f(x^{\alpha}, p^{\alpha})$, whose moments deliver quantities of interest in  relativistic fluid dynamics. The relevant low order moments are the invariant density $h$, the particle flow $N^{\alpha}$, the energy-momentum tensor $T^{\alpha \beta}$ and the energy-momentum flux $M^{\alpha \beta \gamma}$ given by:
\begin{align}\label{eq:moments}
h &= c \int f \frac{d \mybf{p}}{p^0} \;, \\
N^{\alpha} &= c \int f p^{\alpha} \frac{d \mybf{p}}{p^0} \;,\\
T^{\alpha \beta} &= c \int f p^{\alpha} p^{\beta} \frac{d \mybf{p}}{p^0} \;,\\
M^{\alpha \beta \gamma} &= c \int f p^{\alpha} p^{\beta} p^{\gamma} \frac{d \mybf{p}}{p^0} \ .
\end{align}
The transport equation~\cref{eq:vlasov-eq} implies conservation of mass, momentum and energy, expressed here as conservation equations for $N^{\alpha}$, $T^{\alpha\beta}$ and $M^{\alpha\beta\gamma}$~\cite{cercignani-2002,rezzolla-2013,schroeder-2010}:
\begin{align}
\label{eq:cons_eqs1}
0 &= \partial_\alpha N^{\alpha}                                           \; , \\
\label{eq:cons_eqs2}
0 &= \partial_\alpha T^{\alpha\beta} + \frac{e}{c} F^{\beta\alpha} N_{\alpha} \; , \\
\label{eq:cons_eqs3}
0 &= \partial_\alpha M^{\alpha\beta\gamma}+\frac{e}{c}(F^{\beta\alpha}T_{\alpha}^{~\gamma}+F^{\gamma\alpha}T_{\alpha}^{~\beta})                       \; . 
\end{align}
Finally, the electromagnetic field tensor $F^{\alpha\beta}$ (whose components are the electric and magnetic fields, $\mybf{E}$ and $\mybf{B}$, respectively) appearing in~\cref{eq:vlasov-eq} evolves according to Maxwell equations~\cite{jackson-1998}:
\begin{align}
    \label{eq:maxwell-inhomogeneus}
    0 &= \partial_\alpha F^{\alpha\beta} + \mu_0 c e (N^{\beta} + N^{\beta}_{b})\;,  \\
    \label{eq:maxwell-homogeneus}
    0 &= \partial_\alpha F_{\beta\gamma} + \partial_\beta F_{\gamma\alpha} + \partial_\gamma F_{\alpha\beta} \;,  
\end{align}
with $N^{\beta}_{b}$ representing the driving electron bunch that perturbs the plasma.

\subsection{Local Equilibrium Closure (LEC)}\label{sec:lec}
A popular closure for the fluid equations~\cref{eq:cons_eqs1,eq:cons_eqs2,eq:cons_eqs3} is given by the local equilibrium closure (LEC), which postulates the distribution function $f$ solving for~\cref{eq:vlasov-eq} as the Maxwell-J{\"u}ttner distribution~\cite{juettner-1911} (see Appendix~\ref{sec:Appendix2} for details). This hypothesis has some strong implications. First, in this configuration, the fluid must be considered ideal, and therefore its particle flow $N^{\alpha}$ and energy-momentum tensor $T^{\alpha\beta}$ assume the form 
\begin{align}\label{eq:ideal-tensors-fluid}
         N^{\alpha} = n_0 U^{\alpha} \; , \;
    T^{\alpha\beta} = (P_0+\varepsilon_0) \frac{U^\alpha U^\beta}{c^2} - P_0 \eta^{\alpha\beta} \; ,
\end{align}
with $n_0=n/\gamma$, $P_0$, $\varepsilon_0$ respectively the rest number density, pressure, and internal energy density, $U^{\alpha}=\gamma(c,\mybf{u})$ the four vector for fluid velocity ($\gamma$ being its related Lorentz factor), and $\eta^{\alpha\beta}$ is the Minkowsky metric tensor. 
Second, the adoption of a local Maxwell-J{\"u}ttner distribution grants both entropy conservation and the use of an ideal equation of state (in this relativistic framework the Synge~\cite{synge-1957} equation of state, that we adopt here in the small temperature limit). This translates into a particular scaling for temperature $T$, pressure and energy density in the plasma: 
\begin{align}
    \label{eq:lec-temp-scaling}
    T      &= T_i \left(\frac{n_0}{n_i}\right)^{2/3} \; , \\
    \label{eq:lec-pres-scaling}
    P_0    &= n_0 m_e c^2 \mu_i \left(\frac{n_0}{n_i}\right)^{2/3}  \; , \\
    \label{eq:lec-energy-scaling}
\varepsilon_0 &= n_0 m_e c^2 \left[ 1 + \frac{3}{2} \mu_i \left(\frac{n_0}{n_i}\right)^{2/3} \right] \; , 
\end{align}
with $n_i$ and $T_i$ respectively the initial number density and temperature fields, and $\mu_i=\frac{k_b T_i}{m_e c^2}$. All in all, the LEC produces the following set of warm plasma equations: 
\begin{align}\label{eq:lb-ade-lec}
     \partial_t \mybf{A} + \nabla_{\mybf{x}} \cdot (\mybf{u} \mybf{A}) = \mybf{F}        \;,  
\end{align}
where
\begin{align} \label{eq:lb-ade-lec-comps-A}
    \mybf{A} &= 
    \begin{pmatrix}
        n               \\
        \left[ 1 + \frac{5}{2} \mu_i \left(\frac{n_0}{n_i}\right)^{2/3}  \right] n m_e \gamma \mybf{u}
    \end{pmatrix}   \; , 
\end{align}
\begin{align} \label{eq:lb-ade-lec-comps-F}
    \mybf{F} &= 
    \begin{pmatrix}
        0               \\
        - n_i m_e c^2 \mu_i \nabla_{\mybf{x}} \left(\frac{n_0}{ n_i}\right)^{5/3} - e n (\mybf{E} + \mybf{u} \wedge \mybf{B})
    \end{pmatrix}   \; .    
\end{align}
In cylindrical coordinates, assuming axial symmetry and 
imposing no azimuthal fluid velocity $u_{\varphi}=0$ due to the fact that this component is often sub-dominant in these kinds of setups,~\cref{eq:lb-ade-lec} simplifies into a set of three equations, one for the plasma number density $n$ and two for the remaining non-zero components of the relativistic fluid momentum, $\gamma m_e u_r$ and $\gamma m_e u_z$.

\subsection{Warm Plasma Closure (WARMC)} \label{sec:warmc}
A second closure to warm fluid equations is provided by the so called Warm Closure (WARMC)~\cite{schroeder-2010}.~\cref{eq:cons_eqs1,eq:cons_eqs2,eq:cons_eqs3} are re-expressed through the \textit{centered} (w.r.t. the \textit{thermal momentum} $w^{\mu} = N^{\mu}/h$) moments of second ($\theta^{\mu\nu}$) and third order ($Q^{\mu\nu\lambda}$) 
\begin{align}
  \label{eq:theta-integrals}
  \theta^{\mu\nu} &= c \int f (p^{\mu}-w^{\mu}) (p^{\nu}-w^{\nu}) \frac{d \mybf{p}}{p^0} \;,\\
Q^{\mu\nu\lambda} &= 
c \int f (p^{\mu}-w^{\mu}) (p^{\nu}-w^{\nu}) (p^{\lambda}-w^{\lambda}) \frac{d \mybf{p}}{p^0} \;,
\end{align}
where one can easily verify that $\theta^{\mu\nu} = T^{\mu\nu} - \frac{N^{\mu}N^{\nu}}{h}$. The closure is indeed performed by neglecting $Q^{\mu\nu\lambda}$ in the conservation equations, on the excuse of small thermal spreads in the plasma. This leads to the following set of equations 
\begin{align}\label{eq:lb-ade-warmc}
     \partial_t \mybf{A} + \nabla_{\mybf{x}} \cdot (\mybf{u} \mybf{A}) = \mybf{F}        \;,  
\end{align}
where
\begin{align} \label{eq:lb-ade-warmc-comps-A}
    \mybf{A} &= 
    \begin{pmatrix}
        n                                   \\[0.4em]
        n \frac{n_0 U^{\beta}}{h}           \\[0.4em]
        n \frac{\theta^{\beta\gamma}}{h}
    \end{pmatrix}   \; ,
\end{align}
\begin{align} \label{eq:lb-ade-warmc-comps-F}
    \mybf{F} &= 
    \begin{pmatrix}
        0                                                                                   \\[0.4em]
        - \partial_\alpha \theta^{\alpha\beta} - \frac{n_0 e}{c}F^{\beta\alpha}U_{\alpha}   \\[0.4em]
        {\scriptstyle - \theta^{\gamma\alpha} \partial_\alpha \left( \frac{n_0 U^{\beta}}{h} \right) 
        - \theta^{\alpha \beta} \partial_\alpha \left( \frac{n_0 U^{\gamma}}{h} \right) 
        - \frac{e}{c}(F^{\beta\alpha}\theta_{\alpha}^{~\gamma}+F^{\gamma\alpha}\theta_{\alpha}^{~\beta})}
    \end{pmatrix}  \; ,    
\end{align}
which have to be coupled with the constraints coming from the mass-shell condition for kinetic momenta, $p^{\mu}p_{\mu}=m_e^2 c^2$:
\begin{align}
    \label{eq:mass-shell1}
    \theta^\mu_{~\mu} &= h c^2 \left[ m_e^2 - \left(\frac{n_0}{h}\right)^2 \right] \; , \\
    \label{eq:mass-shell2}
    U_\nu \theta^{\mu\nu} &= 0                 \; .
\end{align}
All in all, when considering~\cref{eq:lb-ade-warmc} in a cylindrical geometry with axial symmetry and no azimuthal velocity, this system reduces to a set of eight forced advection equations for the plasma number density $n$, the invariant density $h$, the non-zero cylindrical components of the fluid velocity $u_r$ and $u_z$, and some components of the centered energy momentum tensor, namely $\theta^{rr}$, $\theta^{\varphi\varphi}$, $\theta^{zz}$ and $\theta^{rz}$ (all other non-zero components are recovered via~\cref{eq:mass-shell1,eq:mass-shell2}). \\
Note here that in the WARMC equations, the components of the centered energy-momentum tensor $\theta^{\mu\nu}$, which basically represent pressure and energy density in a generic lab frame of reference (and can indeed be linked to the rest frame versions of said quantities - more details in Appendix~\ref{sec:Appendix1}) appear as dynamic variables, and in principle lead to anisotropic pressures. In PWFA, where processes occur on short time scales so that particle collisions can not act as isotropy-restoring terms, such condition is expected~\cite{shadwick-2004,shadwick-2005}. 
\section{Setup and numerical simulations}\label{sec:setup}
%
\begin{figure}
    \centering
    \includegraphics[width=\columnwidth]{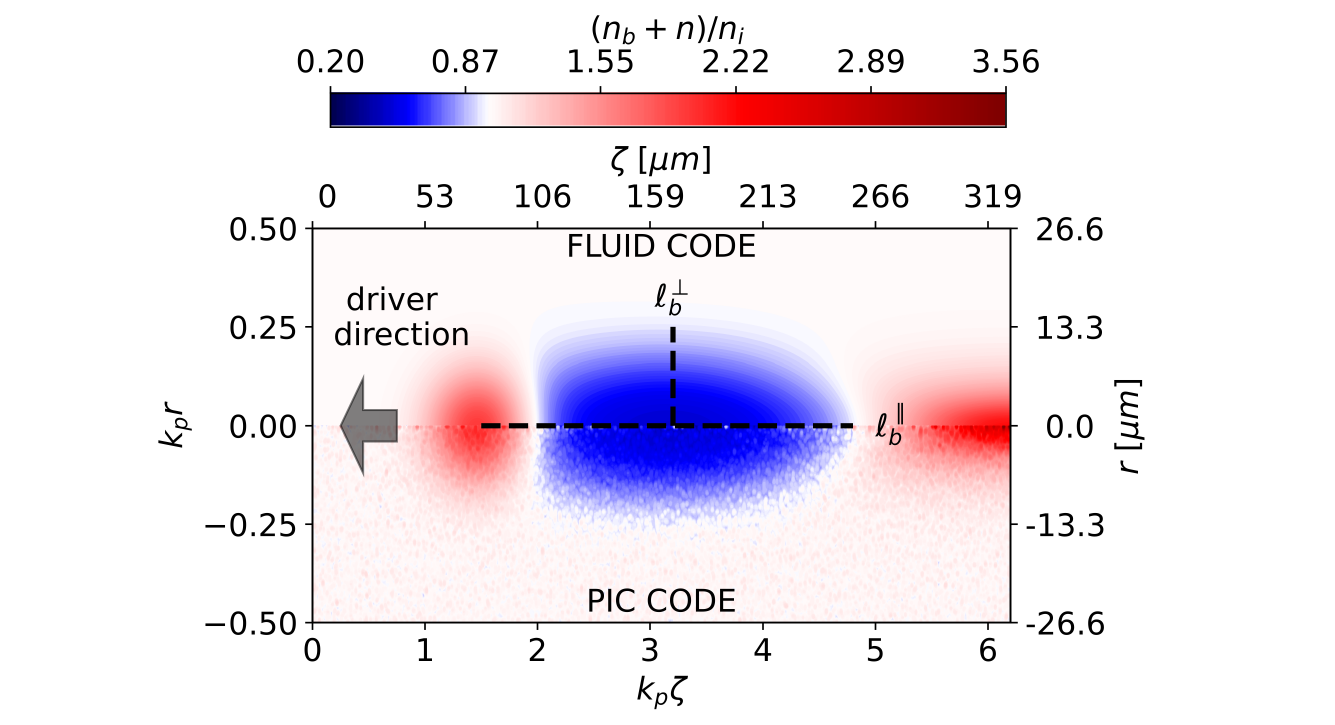}
    \caption{A typical outcome of a fluid solver (top) versus the PIC solver (bottom). Depicted in the figure is the number density, normalized w.r.t. initial unperturbed plasma density $n_i$, given by two contributions: one coming from the Gaussian driving bunch $n_b$, with rms-sizes $\sigma_r = 5.3 \; \rm{\mu m}$ and $\sigma_z = 16 \; \rm{\mu m}$, peak amplitude $\alpha = 1.05 \times 10^{16} \; \rm{cm}^{-3}$, and traveling from right to left at the speed of light. The second contribution is given by the background plasma density $n$, initially set in an equilibrium configuration at uniform temperature $k_b T_i = 37 \; \rm{eV}$ and density $n_i = 10^{16} cm^{-3}$}
    \label{fig:0}
\end{figure}
\begin{figure*}
    \centering
    \includegraphics[width=\textwidth]{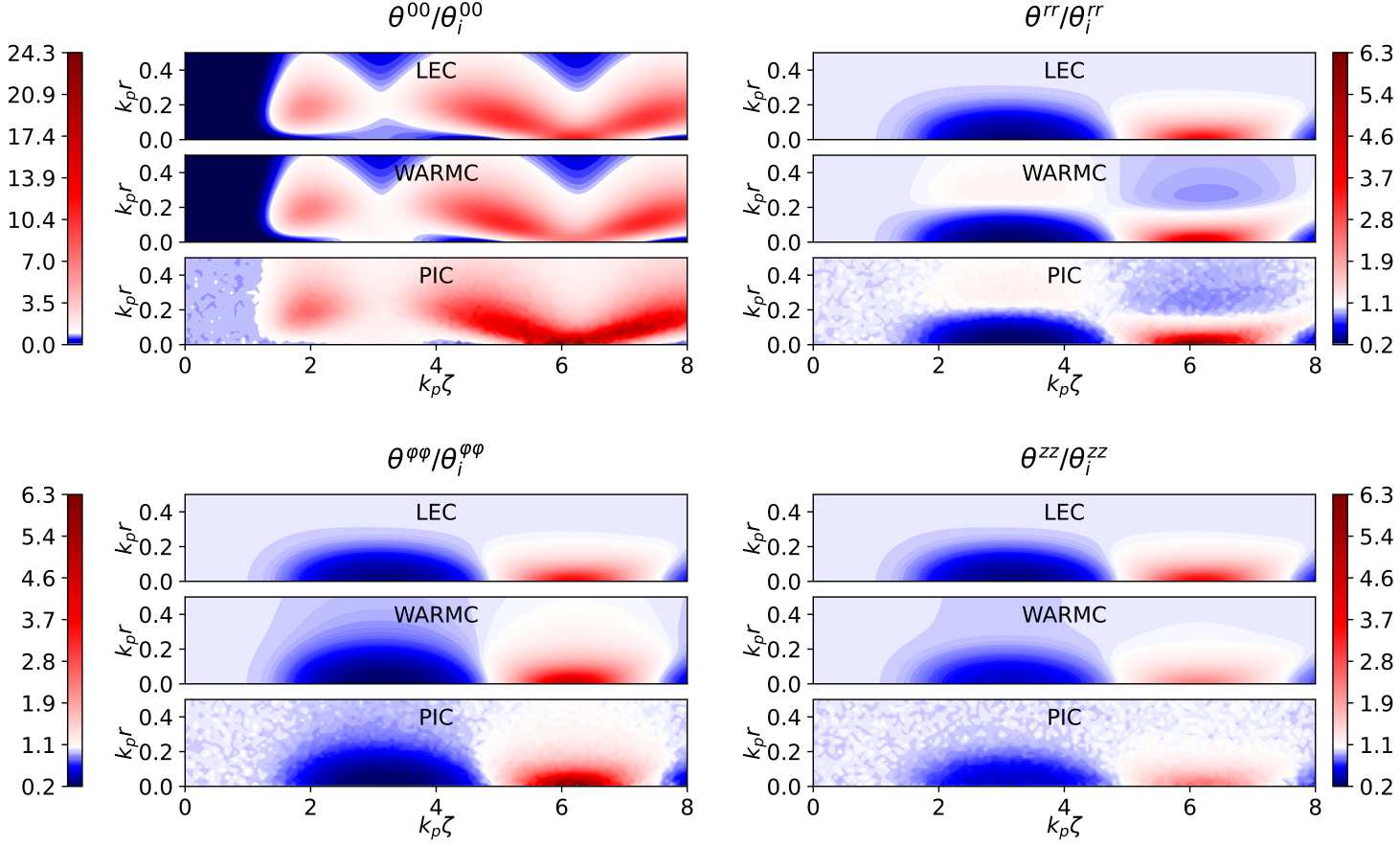}
    \caption{
     Comparison between the results of the fluid code in the two closures (top and middle panels) against the PIC code (bottom panels) for all the diagonal components of the centered energy momentum tensor $\theta^{\mu\nu}$ expressed in cylindrical coordinates. All quantities are normalized w.r.t. their initial values $\theta_i^{00}=n_i 3(k_b T_i)^2/(2m_e c^2)$ and $\theta_i^{rr}=\theta_i^{\varphi\varphi}=\theta_i^{zz}=n_i k_b T_i$, computed in Appendix~\ref{sec:Appendix2} for an isotropic plasma at rest in thermodynamic equilibrium. }
    \label{fig:1}
\end{figure*}
In this section we present the setups employed for the numerical simulations of the fluid equations previously presented, for both the two closure schemes (LEC and WARMC). Such equations can be numerically solved by recurring to the lattice Boltzmann method, with algorithmic details provided in Refs.~\cite{parise-2022, simeoni-2024}. Conversely, Maxwell equations are solved by employing a Finite Difference Time Domain scheme~\cite{yee-1966,olakangil-2000}.
The method has been presented in detail in Refs.~\cite{simeoni-2024, parise-2022}, and the interested reader might refer to these publications for further algorithmic details. \\
In each subsequent figure, we compare results coming from the fluid solver against PIC simulations performed using the code FBPIC~\cite{lehe-2016}. FBPIC is a quasi-3D PIC that employs the Hankel transform to decompose transversely the relevant quantities (the maximum order is set by the user) allowing to go beyond the limiting pure cylindrical symmetry. It also makes use of a spectral solver that does not introduce numerical dispersion for close to the speed of light moving quantities. 
For both the two solvers, we initialize the background plasma at rest, with uniform density $n_i=10^{16} \; \rm{cm}^{-3}$ and temperature $k_b T_i = 37 \; \rm{eV}$: in the case of PIC, such thermal spread is introduced as a variance in the initial velocity distribution of the plasma particles. The plasma is then perturbed by a driver of electrons, with Gaussian bunch density $n_b$
\begin{align}\label{eq:bunch-density}
    n_b = \alpha \, \exp \left(-\frac{(z-z_0)^2}{2\sigma_z^2} - \frac{r^2}{2 \sigma_r^2} \right) \; ,
\end{align}
with $\sigma_r = 5.3 \; \rm{\mu m}$ and $\sigma_z = 16 \; \rm{\mu m}$, and peak amplitude $\alpha = 1.05 \times 10^{16} \; \rm{cm}^{-3}$, set to reproduce the desired value of $0.05$ for the \textit{normalized charge parameter} $\tilde{Q}$, defined as follows:
\begin{align}
    \tilde{Q} = \alpha (2\pi)^{3/2} \sigma_r^2 \sigma_z \left( \frac{k_p^3}{n_i} \right) \; ,
\end{align}
where $k_p = \frac{\omega_p}{c} = \sqrt{e^2 n_i / (m_e c^2 \epsilon_0)}$ is the cold plasma wave-number, $\epsilon_0$ being  vacuum's permittivity. For the fluid solver, the driver is initialized at a position $k_p z_0 = 1.5$ inside the simulation domain, and then moves rigidly at the speed of light from right to left along the z-axis. In the case of the PIC solver, the driving bunch is initialized in vacuum and focused at the vacuum-plasma transition (consisting of a step-like profile) to the prescribed rms-sizes. Its energy is set to $10 \; GeV$, in order to increase the betatron wavelength and reduce any contribution due to transverse evolution. \\
For the fluid code, the simulation domain is $425 \; \rm{\mu m}$ long and has a maximum radius of $318 \; \rm{\mu m}$, with corresponding cell resolutions $k_p dz = k_p  dr = 10^{-2}$. For the PIC code we have instead a box of size $530 \; \rm{\mu m} \times 106 \; \rm{\mu m}$, sampled at a resolution of $k_p dz = 6.2 \times 10^{-3}$ and $k_p dr = 4.7 \times 10^{-3}$. For both solvers, the temporal step is $\omega_ p dt = 10^{-3}$. Finally, in the PIC code, the unperturbed plasma density $n_i$ is sampled with 64 particles per cell ($4 \times 4 \times 4$ in $z,r,\varphi$ directions respectively). This number of particles per cell has been selected after performing a convergence test and chosen as a balanced compromise between computational efficiency and achieving sufficiently smooth fields. In order to reproduce the cylindrical symmetry of the fluid code, only the fundamental azimuthal mode has been employed. \\
In Fig.~\ref{fig:0}, we show a comparison between the fluid and PIC codes at time $1.78 \; \rm{ps}$. We show the total number density, comprising the driving bunch contribution $n_b$ and the plasma response $n$, with the co-moving variable $\zeta = z - c t$ on the x-axis (as all figures presented in this work). This figure both serves as a sketch for the simulation setup, and to verify compliance between the two solvers. \\
Finally, we mention that all the results shown in Sec.~\ref{sec:results} are drawn for the PIC solver by coarse-graining information from the particles of the simulation. In practice, the components of the centered energy-momentum tensor $\theta^{\mu\nu}$, given by~\cref{eq:theta-integrals}, are computed for the PIC code by performing particles averages on sub-cells of the physical domain. Particular care has to be taken when performing such operation, as the sub-cells must have dimensions smaller than the typical length scales of the system, while also being sufficiently large to accumulate adequate statistics for the particles averages. In our case, we employ sub-cells of dimensions $k_p \Delta r = 2 \times 10^{-2}$ and $k_p \Delta z = 6 \times 10^{-2}$. Additionally, when in Sec.~\ref{sec:results} we will be showing slice plots of the different quantities at fixed $r$ values, it is to be intended that such quantities are averaged along the $r$ direction on a slice of size $5 \; \rm{\mu m} = 0.10/k_p$, in order to reduce the noise of the PIC code. 
\section{Results}\label{sec:results}
We present here the results of the comparisons between the fluid solver for both the two fluid closures and the PIC solver, having set up both the two numerical schemes with the prescriptions provided in Sec;~\ref{sec:setup}. First, we compare in Fig.~\ref{fig:1} the diagonal components of the centered energy-momentum tensor: $\theta^{00}$, $\theta^{rr}$, $\theta^{\varphi\varphi}$, $\theta^{zz}$, normalized w.r.t. their initial value at rest (more details on these values at the end of Appendix~\ref{sec:Appendix2}). As it is possible to appreciate in Appendix~\ref{sec:Appendix1}, these are in fact the building blocks used to compute the fluid's rest frame longitudinal and transversal pressure components, and therefore provide already an important indication on matches/mismatches between the fluid schemes and the PIC code. We reckon that, while in the WARMC these components are a direct output of the simulation, and in the LEC case, they have to be computed from rest frame quantities according to~\cref{eq:theta-for-lec}. Finally, as it has already been mentioned, in the PIC case we compute these quantities by performing particle averages of the form~\cref{eq:theta-integrals}. By close inspection of the figure, it is possible to appreciate that the WARMC is in tight visual agreement with the PIC results, and well reproduces all the main spatial features of the tensor components. In particular, it is already possible to appreciate a difference in the spatial entries $\theta^{rr}$, $\theta^{\varphi\varphi}$, $\theta^{zz}$. This inhomogeneity in the spatial components is not expected in the LEC case, where the only differences are due to relativistic effects: in fact in the non-relativistic case these values reduce to the components of Euler-Cauchy stress-tensor of an ideal fluid, see Appendix~\ref{sec:Appendix2} for details. Finally, we mention that at variance with the cited spatial entries of $\theta^{\mu\nu}$, the time-time component $\theta^{00}$'s initial value is quadratic with temperature $T_i$, and hence both the two fluid closures, which are first order theories in $T$, are not expected to recover this quantity as well as the others, both at rest and in the dynamics. Nevertheless, we appreciate a congruent agreement between the PIC and the fluid solver, with most of the spatial features of $\theta^{00}$ well reproduced in the simulations.
\begin{figure*}
    \centering
    \includegraphics[width=\textwidth]{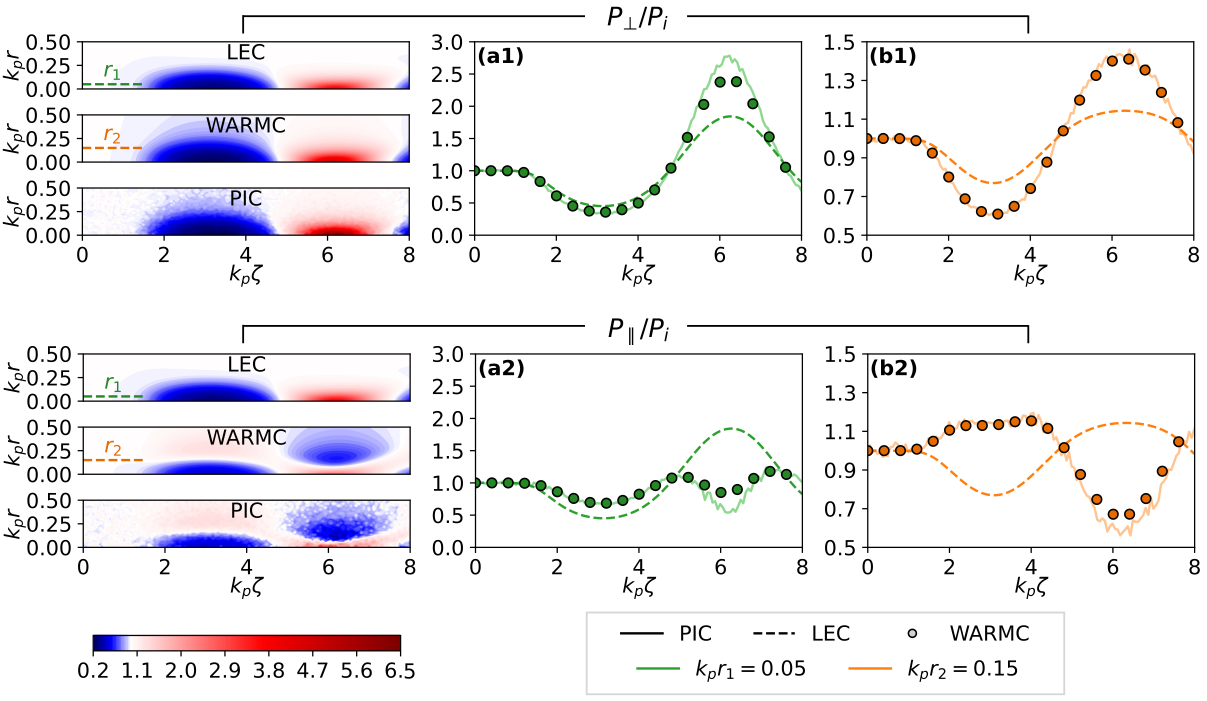}
    \caption{Left column: color plots of the transversal $P_\perp$ (top part of the figure) and longitudinal $P_\parallel$ (bottom part of the figure) rest frame pressure fields for the two fluid theories (top panels LEC, mid panels WARMC) versus the PIC solver (bottom panels). For the LEC fluid theory, these quantities coincide and are computed according to~\cref{eq:lec-pres-scaling}. Middle and right columns: we show two characteristics slice plots at selected values of the $r$ coordinate, $k_p r_1 = 0.05$ (middle column, panels a1-a2) and $k_p r_2 = 0.15$ (right column, panels b1-b2). For all these plots, pressures are normalized w.r.t. their initial isotropic value, $P_i=n_i k_b T_i$.}
    \label{fig:2}
\end{figure*}
\begin{figure*}
    \centering
\includegraphics[width=\textwidth]{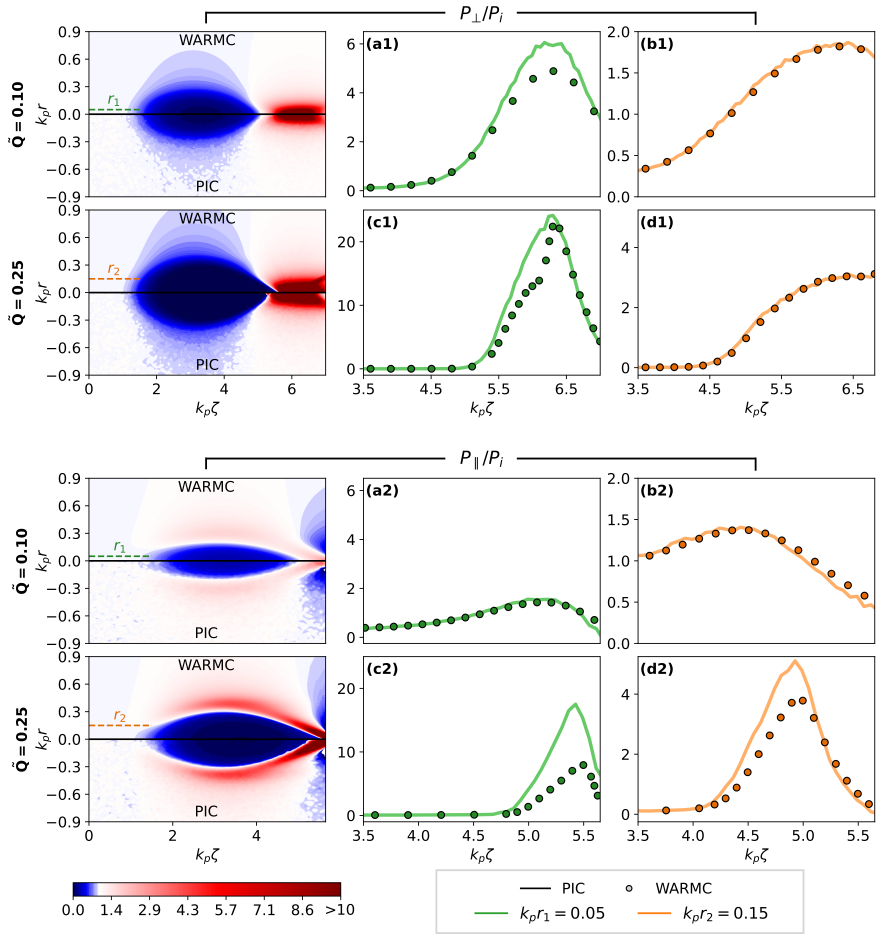}
    \caption{Comparison between the fluid WARMC and the PIC code, for both the transversal (top part of the figure) and longitudinal pressures (bottom part of the figure). We use the setup exposed in Sec.~\ref{sec:setup}, with initial temperature $k_b T_i = 37 \; \rm{eV}$, and for two different values of $\tilde{Q}$: $\tilde{Q}=0.1$ in the first and third row, $\tilde{Q}=0.25$ in the second and fourth row. For both these two parameters, we show color plots (on the left column) and two slice plots at chosen radial values $k_p r_1 = 0.05$ (panels a1-c1-a2-c2) and $k_p r_2 = 0.15$ (panels b1-d1-b2-d2). All pressure components are normalized w.r.t. their initial isotropic value, $P_i = n_i k_b T_i$.}
    \label{fig:3}
\end{figure*}
\begin{figure*}
    \centering
    \includegraphics[width=\textwidth]{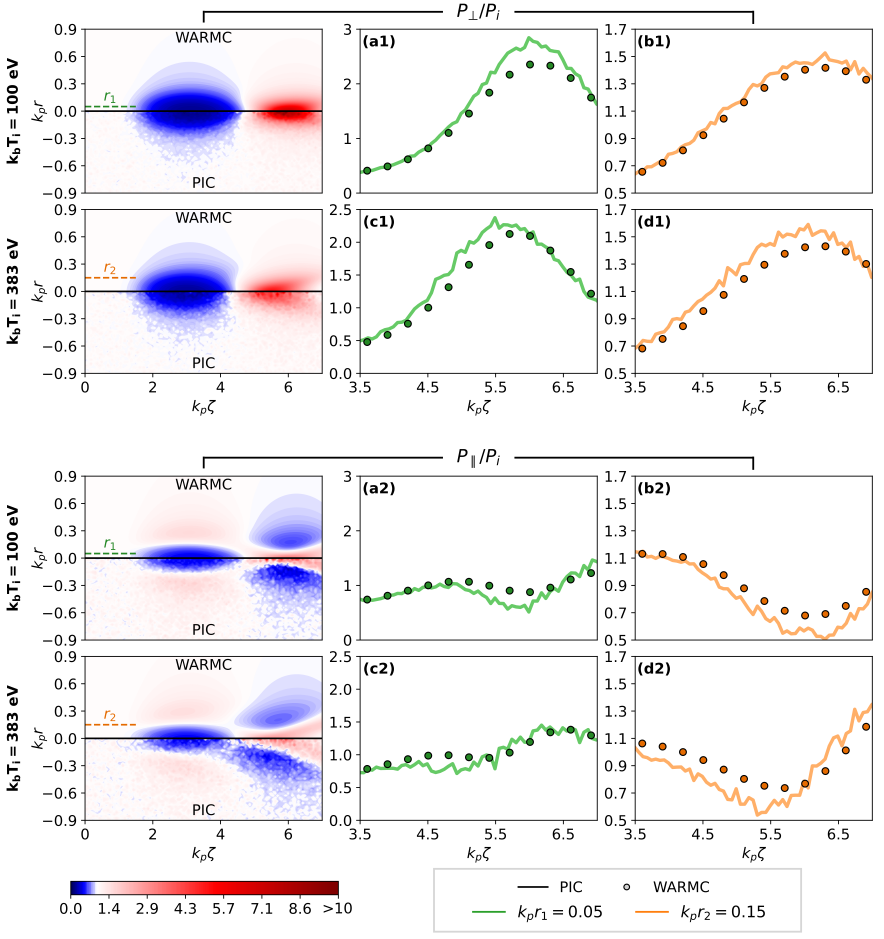}
    \caption{Comparison between the fluid WARMC and the PIC code, for both the transversal (top part of the figure) and longitudinal pressures (bottom part of the figure). We use the setup exposed in Sec.~\ref{sec:setup}, with normalized charge parameter $\tilde{Q}=0.05$, and for two different values of $k_b T_i$: ${k_b T_i} = 100 \; \rm{eV}$ in the first and third row, ${k_b T_i} = 383 \; \rm{eV}$ in the second and fourth row. For both these two parameters, we show color plots (on the left column) and two slice plots at chosen radial values $k_p r_1 = 0.05$ (panels a1-c1-a2-c2) and $k_p r_2 = 0.15$ (panels b1-d1-b2-d2). All pressure components are normalized w.r.t. their initial isotropic value, $P_i = n_i k_b T_i$.}
    \label{fig:4}
\end{figure*}

Evaluating thermal spread anisotropies by comparing components of the centered energy momentum tensor is indeed useful, but a more straightforward characterization can be proposed by recurring to comparisons of the pressure values in the rest frame of the fluid, which basically reduces the parameter space to only two observables (one in the LEC case) instead of many more. As already stated, particle collisions are a restoring mechanism for recovering local equilibrium, and hence pressure isotropy; therefore in the LEC case, where equilibrium is enforced by construction, there is only one single value for pressure, which is constructed from the number density by a simple consequence of ideal gas' law and entropy conservation (\cref{eq:lec-temp-scaling,eq:lec-pres-scaling}). In the WARMC, such constrain is not imposed, and pressure anisotropies might naturally arise in the system. This indeed is observed in Fig.~\ref{fig:2}, where we compare the isotropic pressure term coming from LEC against the longitudinal $P_\parallel$ and transversal $P_\perp$ pressure terms that can be derived in the WARMC case: these values are indeed obtained connecting the numerically computed components of the centered energy-momentum tensor in the lab frame, $\theta^{\mu\nu}$, to its rest fluid's frame counterpart, via a Lorentz transformation, see Appendix~\ref{sec:Appendix1} for more details. When comparing against the PIC, we see, at variance with the LEC model, that there is a very good agreement with the WARMC theory, both qualitatively (color plots on the left) and quantitatively (slice plots on the right). This is especially true in the first electron depletion bubble after the driver, and the agreement is generally well maintained all over the spatial domain. Close to the radial axis (see for example Fig.~\ref{fig:2}(a1-a2)), approximately at the closing of the first electron depletion bubble in $k_p \zeta \sim 5-6$, we observe a tenuous mismatch: this is to be expected, since judging by Fig.~\ref{fig:0} this is where the density peak is approximately located, and hence where the agreement between fluid theories and kinetic solvers is expected to be worsening. In addition, the particle averages for the coarse-graining of the PIC quantities (see Sec.~\ref{sec:setup}) are most sensible to sharp field transitions, and this is exactly the case in this particular zone. Nevertheless, by comparing the color plots and slice plots for both $P_{\perp}$ (see Fig.~\ref{fig:2}(a1-b1)) and $P_{||}$ (see Fig.~\ref{fig:2}(a2-b2)) we see a convincing indication of the occurrence of thermal spread anisotropies in the plasma, i.e. $P_{||} \neq P_{\perp}$; additionally, we note that that the biggest differences w.r.t. the LEC isotropic case are encountered for $P_{||}$: this means that the LEC would fail to reproduce longitudinal pressure forces while still being able to qualitatively reproduce radial expansion/compression due to thermal effects. This feature is probably due to the relativistic nature of the driving bunch; in relativistic beam dynamics, the beam temperature typically have different values (and may also have different definitions) in the longitudinal and transverse directions (see, for example, Ref.~\cite{reiser-2008-338}); in addition, a number of collisional effects display similar non isotropic behaviors like the Boersch~\cite{reiser-2008-472} or the Touschek~\cite{reiser-2008-482} effects. It does seem then reasonable to attribute the failure of LEC on the longitudinal direction to the boosting of the driver reference frame while in the transverse direction it retains some validity due to the relative "smallness" of the perturbation occurring in the linear regime. One last remark is in place: this figure provides indication that a selection of the closure scheme might impact on the radial size of the first depletion bubble: from the slice plots it is indeed possible to appreciate that the way the radial pressure is restored to the initial pressure is different in the LEC and WARMC. Estimates on the radial size of the first depletion bubble in the wake of the driver are indeed a topic of interest in PWFA~\cite{lu-2006,tzoufras-2008,dalichaouch-2021,golovanov-2023}, and we reserve a characterization study of the dependence of these quantities on thermal spread to future works. \\
Up to now, we have adopted for our comparisons a fixed setup, comprising an initially uniform temperature field of $k_b T_i = 37 \; \rm{eV}$ and a normalized charge parameter of $\tilde{Q} = 0.05$. The question of how this characterization stands up to variations of these parameters might naturally come and has indeed to be faced. On one side, fluid theories cannot sustain highly non-linear regimes, as they are expected to break down in the presence of the strong field gradients that might arise in such situations. Conversely, warm fluid closure schemes are formulated on the assumption of small thermal spreads, and hence are expected to deliver incorrect results at rising values of $k_b T_i$.  We therefore characterize the impact of an increase in non-linearity and an increase in temperature in the system. Given the very good agreement that we have found between PIC simulations and WARMC (see Fig.~\ref{fig:2}), we restrict our analysis to the WARMC. In Fig.~\ref{fig:3} we show pressure fields by fixing the initial temperature to $k_b T_i = 37 \; \rm{eV}$ and increasing the level of non-linearity of the system. We choose two values of the normalized charge parameter $\tilde{Q}=0.1$ and $\tilde{Q}=0.25$, and regarding the slice plots we focus on the $\zeta$ range on the most problematic area for the matching, i.e., the tail of the electron depletion bubble, as the area at the start of the bubble is clearly caught by the WARMC. 
By increasing the non-linearity in the system, we actually observe that, although not dramatically, the WARMC-PIC matching starts to lose consistency and presents a mismatch for both $P_{\perp}$ (see Fig.~\ref{fig:3}(a1-b1) versus Fig.~\ref{fig:3}(c1-d1)) and $P_{||}$ 
(see Fig.~\ref{fig:3}(a2-b2) versus Fig.~\ref{fig:3}(c2-d2)). The mismatch starts around the density peak, and moves to the left toward the bulk of the depletion bubble. For $P_{\perp}$ the mismatch is particularly evident close to the axis (Fig.~\ref{fig:3}(a1)-(c1)) and not much pronounced away from the axis (Fig.~\ref{fig:3}(b1)-(d1)); for $P_{||}$ we find a mismatch both close to the axis (Fig.~\ref{fig:3}(a2)-(c2)) and away from it (Fig.~\ref{fig:3}(b2)-(d2)). It is worth noting that, in the $\tilde{Q}=0.25$ case, the longitudinal size of the bubble starts diverging between the two models. This hints at a clear dependency of this observable on the chosen closure for the fluid equations, and could be worth future research. Conversely, we move our analysis (Fig.~\ref{fig:4}) to a comparison of pressure fields for increasing values of the initial temperature, which we fix at $k_b T_i = 100 \; \rm{eV}$ and $k_b T_i = 383 \; \rm{eV}$. We choose the normalized charge parameter to be $\tilde{Q} = 0.05$, and keep the same setup exposed in Sec.~\ref{sec:setup}. In this case, there is again the onset of a divergence between the PIC and WARMC results, although we observe that the divergence at increasing temperature is not as dramatic as the one observed at increasing non-linearity. Nevertheless, despite the increase in noise at raising $k_b T_i$, it is observed again that the mismatch starts from the density peak and moves backward toward the start of the bubble  (see Fig.~\ref{fig:4}(c1-d1) and relative color plot). For the higher $k_b T_i$ value, there is also a clear mismatch in the estimation of the longitudinal wake features. This suggests once again further investigations on the impact of finite temperatures and chosen fluid closures on the spatial features of the wakefield in PWFA experiments. \\
All in all, we infer from both Fig.~\ref{fig:3} and Fig.~\ref{fig:4} that for the chosen parameters the WARMC is still capable of reproducing qualitatively the dynamic shown by the PIC. Nevertheless, by increasing $\tilde{Q}$ and $k_b T_i$ one might start diverging importantly from the kinetic solution, and should be careful to adopt the WARMC for the evaluation of PWFA systems.

\section{Conclusions}\label{sec:conclusions} 
We have used spatially resolved warm fluid models to characterize thermal spread anisotropies in a typical setup of plasma wakefield acceleration (PWFA), where a neutral plasma with finite temperature is perturbed by a bunch of relativistic particles (driver). Our focus was on the initial stages of the perturbation induced by the driver, where the dynamics take place on time-scales where local equilibration of the background plasma is not granted. This poses the question on what is the right closure scheme to be applied to the moment equations derived from the relativistic Vlasov equation to formulate the warm fluid model. In this landscape, pressure anisotropies are the hallmark pointing to the correctness of the adopted closure scheme. We have therefore evaluated pressure anisotropy effects in warm fluid models based on different closure assumptions and compared with particle-in-cell (PIC) simulations. We found that there is a window of applicability for warm fluid models, with a clear evidence that fluid closures based on higher order truncations of the kinetic moments equations (WARMC)~\cite{schroeder-2005,schroeder-2009,schroeder-2010} are more prone to reproduce a faithful dynamics for the plasma.\\ 
Various pathways for future research can be envisaged. Warm fluid models are not supposed to work for large temperatures and strongly non-linear wakes, and a detailed characterization of the limiting temperature and degree of non-linearity must also depend on the structure of the driver bunch, a feature that we have not explored in detail in our analysis. Moreover, a dedicated and complementary analysis focused on the characterization of density, electric and magnetic fields and velocity in the wake of the driver would also be needed on a future perspective to better clarify the validity and limitations of the various closure schemes. Finally, we remark that our results provide a clear indication that temperature has a non-trivial effect on the size of the first bubble in the wake of the driver bunch. These effects might be relevant for the PWFA community, due to the interest in the theoretical and numerical evaluation of the boundaries of the first electron cavity in the blowout regime~\cite{lu-2006,tzoufras-2008,dalichaouch-2021,golovanov-2023}, which directly reflects on the intensity of the longitudinal accelerating wakefields. In this landscape, warm fluid models could help in rationalizing how temperature impacts such scenario, thus requiring a thorough study on the effects of fluid closures on the dynamics, particularly on the intensity of the wakefields. 

\section*{Acknowledgments}

The authors gratefully acknowledge Fabio Bonaccorso for his technical support. They also wish to thank Alessandro Cianchi, Alessio Del Dotto and Stefano Romeo for fruitful discussions. This work was supported by the Italian Ministry of University and Research (MUR) under the FARE program (No. R2045J8XAW), project "Smart-HEART". MS gratefully acknowledges the support of the National Center for HPC, Big Data and Quantum Computing, Project No. CN\_00000013 - CUP E83C22003230001, Mission 4 Component 2 Investment 1.4, funded by the European Union - NextGenerationEU. Financial support from the project DYNAFLO (CUP E85F21004290005) of Tor Vergata University of Rome is acknowledged.


\section*{Author Declarations}

\subsection*{Conflict of Interests}
The authors declare that they have no conflict of interests.

\subsection*{Author Contributions}

\textbf{Daniele Simeoni}: Conceptualization (equal); Data Curation (lead); Formal Analysis (lead); Investigation (lead); Software (lead); Visualization (lead); Writing - original draft (equal). \textbf{Andrea Renato Rossi}: Conceptualization (equal); Investigation (supporting); Software (supporting); Writing - original draft (supporting).
\textbf{Gianmarco Parise}: Conceptualization (supporting); Software (supporting); Writing - original draft (supporting).
\textbf{Fabio Guglietta}: Conceptualization (supporting); Software (supporting); Writing - original draft (supporting).   
\textbf{Mauro Sbragaglia}: Conceptualization (equal); Supervision (lead); Writing - original draft (equal).


\section*{Data Availability}

The data that support the findings of this study are available from the corresponding author upon reasonable request.


\appendix

\section{Pressure calculations in the rest frame}\label{sec:Appendix1}

In this section, we will determine the link between quantities (denoted with subscript $\null_0$) defined in the rest frame of the fluid (where $U^{\mu}=U^{\mu}_0=(c,0,0,0)$ and the lab-frame quantities, namely, $N^{\mu}$, $T^{\mu\nu}$ and $\theta^{\mu\nu}$. This derivation is in principle independent of the underlying phase distribution function of the plasma, but we will see in Appendix~\ref{sec:Appendix2} that an assumption of local equilibrium (i.e., a LEC to fluid equations) is indeed compatible with the findings of this section. \\
We define, extending~\cite{cercignani-2002}, an anisotropic energy-momentum tensor $T^{\mu\nu}_0$ in the fluid's rest frame~\cite{letelier-1980}, with pressure $P_\perp$ in the directions transversal to the motion of the driver's bunch, and $P_\parallel=P_\perp+\Delta P$ in the direction longitudinal to its motion (we omit rest-frame subscripts $\null_0$ for these pressures, in order to keep the notation light)
\begin{align}
    T_0^{\alpha\beta} = \text{diag} \left( \varepsilon_0, P_\perp, P_\perp, P_\parallel \right) \; ,
\end{align}
together with the rest frame particle flow $N_0^{\alpha}$
\begin{align}
    N_0^{\alpha} = n_0 (c, 0, 0, 0) \; .
\end{align}
It is then immediate to build $\theta^{\mu\nu}_0 = T^{\mu\nu}_0 - \frac{N^{\mu}_0N^{\nu}_0}{h}$:
\begin{align}\label{eq:theta-0}
    \theta^{\alpha\beta}_0 = \text{diag} \left( \varepsilon_0 - \frac{n_0^2c^2}{h}, P_\perp, P_\perp, P_\perp \right) + \Delta P \delta^{\alpha}_{~z} \delta^{\beta}_{~z} \; .
\end{align}
To get the corresponding quantities in a generic frame of reference, it is then sufficient to apply a Lorentz transformation $\Lambda^{\mu}_{~\alpha}$ 
to the tensor in~\cref{eq:theta-0}, $\theta^{\mu\nu} = \Lambda^{\mu}_{~\alpha} \Lambda^{\nu}_{~\beta} \theta^{\alpha\beta}_{0}$: 
\begin{align}\label{eq:double-boost}
    \theta^{\mu\nu} &= \left( P_\perp + \varepsilon_0 - \frac{n_0^2c^2}{h} \right) \frac{U^\mu U^\nu}{c^2} - P_\perp \eta^{\mu\nu} + \Delta P \Lambda^{\mu}_{~z} \Lambda^{\nu}_{~z} \;. 
\end{align}
Assuming axial symmetry and no azimuthal velocity ($u^{\varphi}=0$), the relevant components of~\cref{eq:double-boost}, properly converted into cylindrical coordinates, can then be inverted to deliver the desired rest frame quantities. For the two pressure values, $P_\parallel$ and $P_\perp$, one gets:
\begin{align}
    P_\perp     &= \theta^{\varphi\varphi} \; ; \\
    P_\parallel &= \frac{c^2(\theta^{rr}+\theta^{zz}-\theta^{\varphi\varphi})-\vert\mybf{u}\vert^2\theta^{00}-u_r^2\theta^{\varphi\varphi}}{c^2+u_z^2}
\end{align}
Finally, we remark that recovering pressure isotropy (as is the case for the LEC), would simply reduce to setting $\Delta P = 0$ and hence $P_\parallel=P_\perp=P_0$ in~\cref{eq:double-boost}. This would deliver a proper link between the centered energy-momentum tensor $\theta^{\mu\nu}$ of an ideal fluid and its rest quantities $\varepsilon_0$, $P_0$ and $h$. Therefore, in the LEC case the most relevant components of $\theta^{\mu\nu}$ would read as:
\begin{align}
    \theta^{00}             &= \left( P_0 + \varepsilon_0 - \frac{n_0^2c^2}{h} \right) \gamma^2 \frac{c^2 + \vert \mybf{u} \vert^2}{c^2} - P_0           
           \:, \\
    \theta^{rr}             &= \left( P_0 + \varepsilon_0 - \frac{n_0^2c^2}{h} \right) \gamma^2 \frac{u_r^2}{c^2}
     + P_0 \:, \\ 
    \theta^{\varphi\varphi} &= P_0                                                                              
           \:, \\
    \theta^{zz}             &= \left( P_0 + \varepsilon_0 - \frac{n_0^2c^2}{h} \right) \gamma^2 \frac{u_z^2}{c^2} 
    + P_0  \; .
\end{align}
In Sec.~\ref{sec:Appendix2}, we present how to obtain these same quantities in the LEC case by direct calculation of the moments of the Maxwell-J{\"u}ttner distribution.
We note here that in the limit of small velocities, these tensors recover the usual form of non-relativistic thermodynamics:
\begin{align}
    \label{eq:theta-00-nonrel}
    \theta^{00}             &= \left(\varepsilon_0 - \frac{n_0^2c^2}{h} \right) 
                            + O \left( \frac{\vert \mybf{u}\vert^2}{c^2} \right)               \:, \\
    \label{eq:theta-rr-nonrel}
    \theta^{rr}             &= P_0 + O \left( \frac{u_r^2}{c^2} \right)             \:, \\
    \label{eq:theta-pp-nonrel}
    \theta^{\varphi\varphi} &= P_0                                                  \:, \\
    \label{eq:theta-zz-nonrel}
    \theta^{zz}             &= P_0 + O \left( \frac{u_z^2}{c^2} \right)             \; .
\end{align}
In particular, the diagonal spatial components of $\theta^{\alpha\beta}$ are all equal and reduce to the usual components of the Euler-Cauchy stress tensor. \\
In Sec.~\ref{sec:Appendix2} we present how to obtain these same quantities in the LEC case by direct calculation of the moments of the Maxwell-Jüttner distribution.

\section{LEC integrals}\label{sec:Appendix2}

This section reports results drawn from Ref.~\cite{gabbana-2021} and expand them. The interested reader might refer to this publication for a more in depth discussion. \\
In the LEC, the plasma is assumed to be at equilibrium, i.e., it is assumed that the phase-space distribution function $f$ solving for~\cref{eq:vlasov-eq} is the Maxwell-J{\"u}ttner distribution~\cite{juettner-1911}
\begin{align}\label{eq:mj-dist}
    f^{eq} &= 
    \frac{n_0}{4 \pi m_e^2 c k_b T \kappa_2} 
    exp{\left(-\frac{p^{\mu}U_\mu}{k_b T}\right)} \; ,
\end{align}
where $\kappa_\nu=K_\nu\left( \frac{m_e c^2}{k_b T} \right)$ is the modified Bessel function of the second kind of index $\nu$. One can therefore plug~\cref{eq:mj-dist} into~\cref{eq:moments} to compute the first hydrodynamic moments of $f^{eq}$
\begin{align}\label{eq:computed-moments}
                h &= \frac{n_0}{m_e} \frac{\kappa_1}{\kappa_2}                        \;, \\
       N^{\alpha} &= n_0 U^{\alpha}                                                 \;,\\
 T^{\alpha \beta} &= \left( n_0 m_e c^2 \frac{\kappa_3}{\kappa_2} \right) \frac{U^{\alpha}U^{\beta}}{c^2} 
                   - n_0 k_b T \eta^{\alpha\beta}                                   \;. 
\end{align}
A few comments are in place. As it has been already anticipated, the tensorial structure of the particle flow and the energy momentum tensor is the same given in the main text,~\cref{eq:ideal-tensors-fluid}, and provides in a natural way both the ideal gas law and the Synge Equation of State as
\begin{align}
    P_0             &= n_0 k_b T    \; , \\
    \varepsilon_0   &= n_0 m_e c^2 \frac{\kappa_3}{\kappa_2} - n_0 k_b T \ ,
\end{align}
that reduce to~\cref{eq:lec-pres-scaling,eq:lec-energy-scaling} in the case of small temperatures. Second, from~\cref{eq:computed-moments} it is possible to derive the form of the centered energy-momentum tensor $\theta^{\mu\nu} = T^{\mu\nu} - \frac{N^{\mu}N^{\nu}}{h}$ in the LEC case 
\begin{align}\label{eq:theta-for-lec}
    \theta^{\mu\nu} = \left( P_0 + \varepsilon_0 - \frac{n_0^2c^2}{h} \right) \frac{U^\mu U^\nu}{c^2} - P_0 \eta^{\mu\nu} \;. 
\end{align}
and to realize that this form is exactly the one derived in~\cref{eq:double-boost} with $\Delta P = 0$. Finally, it can be convenient to elaborate a bit on the initial values that are assumed by $\theta^{\alpha\beta}$ in the case of an unperturbed plasma. In these last considerations, valid for both LEC and WARMC, we will assume the following:
\begin{enumerate}
    \item The unperturbed plasma is in equilibrium, i.e., its mesoscopic state is described by the local equilibrium~\cref{eq:mj-dist};
    \item The plasma is uniform and at rest, i.e., it has constant number density $n_0=n=n_i$ and zero velocity $\mybf{u}=0$;
    \item Due to 1), the plasma fluid can be considered to be isotropic, i.e., $\Delta P = 0$;
\end{enumerate}
With these prescriptions, the asymptotic values given in~\cref{eq:theta-00-nonrel,eq:theta-rr-nonrel,eq:theta-pp-nonrel,eq:theta-zz-nonrel} become exact, and we can further elaborate on them by employing the findings of this section. In the small temperatures limit, one gets: 
\begin{align}
    \label{eq:theta-00-init}
    \theta^{00}_i               &= \left( \varepsilon_i - \frac{n_i^2 c^2}{h_i} \right)
                                 = n_i \frac{3(k_b T_i)^2}{2m_e c^2} + O(T_i^3)  \:, \\
    \label{eq:theta-rr-init}
    \theta^{rr}_i               &= P_{i} = n_i k_b T_i\:, \\
    \label{eq:theta-pp-init}
    \theta^{\varphi\varphi}_i   &= P_{i} = n_i k_b T_i\:, \\
    \label{eq:theta-zz-init}
    \theta^{zz}_i               &= P_{i} = n_i k_b T_i\; .
\end{align}
     
\printbibliography

\vfill{}

\begin{center}
\small
This article may be downloaded for personal use only. Any other use requires prior permission of the author and AIP Publishing. This article appeared in~\cite{simeoni-2024-b} and may be found at \url{https://doi.org/10.1063/5.0216707}
\end{center}

\end{document}